\documentclass{epl}

\title{A schlieren method for ultra-low angle light scattering measurements}
\author{Doriano Brogioli \and Alberto Vailati \and Marzio Giglio}
\institute{
Dipartimento di Fisica and Istituto Nazionale per la Fisica della Materia -
Universit\`a degli Studi di Milano, via Celoria 16,
20133 Milano, Italy
}
\shortauthor{D. Brogioli \etal}
\shorttitle{Schlieren method for light scattering}

\begin{document}

\maketitle

\begin{abstract}
We describe a self calibrating optical technique that allows to
perform absolute measurements of scattering cross sections for the
light scattered at extremely small angles. Very good
performances are obtained by using a very simple optical layout
similar to that used for the schlieren method, a technique
traditionally used for 
mapping local refraction index changes. The scattered intensity
distribution is recovered by a statistical analysis of the random
interference of the light scattered in a half-plane of the scattering
wave vectors and the main transmitted beam. High quality data can be
obtained by proper statistical accumulation of scattered intensity
frames, and the static stray light contributions can be eliminated
rigorously. The potentialities of the method are tested in a
scattering experiment from non equilibrium fluctuations 
during a free diffusion experiment. Contributions of light scattered
from length scales as long as $\Lambda=1\un{mm}$ can be accurately
determined.
\end{abstract}

Low angle static light scattering is a powerful tool to investigate the
long wavelength fluctuations that appear in a great variety of important
phenomena, like phase transitions and nucleation processes \cite{stanley,domb},
aggregation of colloidal systems \cite{lin89,asnaghi97},
and non equilibrium systems \cite{vailati97}.
As the typical length scale $\Lambda$ increases, the angular range of
interest $\delta\Theta=\lambda/\Lambda$ becomes smaller, and this creates
unsormountable difficulties stemming from the cramped solid angle of
collection for the scattered light, tighter angular definition for the
scattering angle, ever increasing difficulty with stray light that is
preferentially scattered at smaller angles, and finally main beam
diffraction spilling that also piles up at the low angle scattering end. 

In this letter we describe a new technique that allows to overcome
most of the above mentioned difficulties and can be pushed to
arbitrary small angles. This technique relies on a detection scheme totally
different from the traditional angle-resolved scattering
method \cite{brogioli_PhD}, and based on an optical setup similar to schlieren
imaging technique, traditionally used to map refraction index
modulations \cite{goodman_fourier_optics}.

The technique belongs to a family of scattering
measurement techniques recently introduced, Near Field Scattering
(NFS) \cite{giglio00,giglio01,brogioli02}. These techniques allow to
measure scattered
intensity distributions from the statistical analysis of the intensity
modulations due to the random interference of the scattered waves in
the near field.
In Heterodyne NFS (HNFS) \cite{brogioli02}, the scattered light is
mixed with the 
transmitted beam that acts as a reference beam. The scattered intensity
distribution is then obtained from the two dimensional power spectrum of
the heterodyne signal falling onto a two dimensional multielement sensor (a
CCD). Indeed, as we will stress later on, for small scattering angles
the HNFS becomes a Shadowgraphy  technique
\cite{cannell95,cannell96,brogioli00,brogioli00_2}.
For transparent samples
the transfer function has a wide, pronounced zero around $q=0$
(as well as multiple zeros at higher scattering wavevectors), and this
numbs away the sensitivity where it is mostly sought, namely at extremely
small scattering angles. We will show that one of the prominent
features of Schlieren-like Near Field Scattering (SNFS) is that its
transfer function is flat, irrespective 
of the investigated wave vector range.
We will show that the physical source for the shadow zeros is
thus destroyed, and the HNFS method can be restored to arbitrarily small
scattering wave vectors. 

To show the potentiality of the method, we 
have applied SNFS to the
investigation of non-equilibrium  
fluctuations arising in a free diffusion process in ordinary binary
mixtures \cite{vailati97}.
Nonequilibrium fluctuations scatter light within a very narrow
wavevector range in the forward direction, and the scattered intensity
decay at large scattering wave vectors follows a simple, steep power
law behaviour. Therefore non equilibrium fluctuations are an almost
ideal test sample to ascertain both the linearity of the response and
the $q$ vector range of the method.
Although SALS has been effectively used to investigate the nonequilibrium fluctuations in
near critical binary mixures, attempts to use SALS to investigate the
fluctuations during free diffusion in ordinary binary liquid mixtures
showed that the scattered light is dominated by spurious
contributions, due to the main beam diffraction spilling and to the
divergence of stray light at small scattering angles.  
On the contrary, shadowgraph has been
succesfully used to perform these measurements \cite{brogioli00,brogioli00_2}.
Unfortunately, shadowgraph sensitivity vanishes for small wave
vectors, and also exhibits  
multiple zeroes at higher wave vectors. By contrast, we will show
that SNFS transfer function is a constant as a function of scattering
wave vector, and it significantly extends the wave 
vector range of SALS, shadowgraph and NFS techniques. SNFS allows a
rigorous subtraction of stray light, without any blank measurement;
being a heterodyne 
technique, it measures the field amplitude, thus providing a wider
dynamic range with respect to intensity measurements; also being a
self referencing method, it allows the determination of absolute
differential scattering cross sections.

\begin{figure}
\onefigure{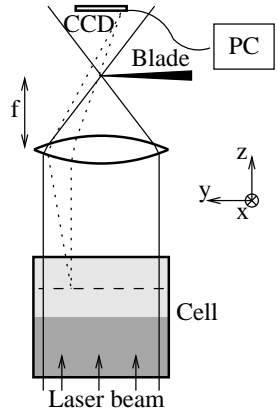}
\caption{
The overall instrumental setup. A collimated laser beam is
sent through the cell containing two miscible liquids. A lens images a
plane inside the cell onto a CCD detector. The transmitted beam is
focused by the lens, and the blade cuts half of its waist.
}
\label{fig_setup}
\end{figure}

The optical layout of the system is shown in fig.~\ref{fig_setup}.
The output of a 10\un{mW} He-Ne laser is spatially filtered and
collimated, so to obtain a 2\un{cm}-diameter beam at $1/e^2$, and sent
through the sample.
Let us call $\vect{k}_0$ the wave vector of the main beam; the goal is to measure
$I_s\left(\vect{Q}\right)$, the intensity of the light scattered at
wave vector $\vect{k}_s$, 
as a function of the transferred momentum $\vect{Q}=\vect{k}_s-\vect{k}_0$.
A lens with focal
length $f=10\un{cm}$ images a plane inside the sample onto a CCD detector
with a magnification about 1 (the exact location of the plane is
immaterial); the images are made of $512\times512$ pixels, and
are digitized with a dynamic range of $8$ bits. 
The transmitted beam is focused by the lens, and
its $3\un{\mu m}$-diameter waist is located between the lens and the
CCD detector. A blade
%
mounted on an x,z micropositioner
%
is placed perpendicular to the optical axis, so that its edge is in
the beam waist.
%
The position of the blade along the optical axis is carefully adjusted
so that the blade does not produce diffraction fringes when it crosses
the beam.
Due to the high sensitivity of the instrument, care has to be taken to
avoid even faint airborne disturbancies. Therefore the entire beam path has
been shielded with cardboard tubing.
%

The coordinate system we will use in the following is defined so that the $z$ axis is  
the optical axis, and the $y$ axis is perpendicular to 
the edge of the blade. Every beam scattered with a negative component
of $\vect{k}_s$ along the $y$ axis is focused onto the blade, and thus
removed. 
Since the scattered field is weak compared to the transmitted beam, 
the intensity $I\left(x,y\right)$ is the sum of the strong transmitted beam $I_0$
and of the small heterodyne term $\delta I\left(x,y\right)$, due to the interference 
between the transmitted and the scattered beams; homodyne terms arising from the 
interference between the scattered beams can be neglected.

The intensity fluctuation $\delta I\left(x,y\right)$ can be decomposed in its Fourier 
components, with amplitude $\delta I\left(q_x,q_y\right)$. A modulation with wave vector
$\vect{q}=\left[q_x,q_y\right]$ is generated by the interference of the transmitted beam 
with the scattered three-dimensional plane wave with wave vector 
$\vect{k}_s\left(\vect{q}\right)=\left[q_x,q_y,k_z\right]$ and amplitude
$\delta E\left(q_x,q_y,k_z\right)$, and therefore $\delta I\left(q_x,q_y\right) \propto 
\delta E\left(q_x,q_y,k_z\right)$. In practice, we evaluate the power spectrum 
$S_{\delta I}\left(\vect{q}\right)$, the mean square value of 
$\delta I\left(q_x,q_y\right)$. This gives the mean intensity of the wave with amplitude 
$\delta E\left(q_x,q_y,k_z\right)$, that is $I_s\left(\vect{Q}\right)$:
\begin{equation}
S_{\delta I}\left(\vect{q}\right)=
I_s\left[\vect{Q}\left(\vect{q}\right)\right],
\label{eq_spettro}
\end{equation}
where $\vect{Q}\left(\vect{q}\right)=\vect{k}_s\left(\vect{q}\right)-\vect{k}_0$.
Consequently, the scattered intensity can be evaluated by measuring the
2-dimensional power spectrum.
Since elastic scattering is considered, $k_z$ is determined by the
condition $\left|\left[q_x,q_y,k_z\right]\right|=k$, with
$k=\left|\vect{k}_0\right|$.
We can thus easily evaluate the transferred wave vector 
$\vect{Q}\left(\vect{q}\right)=\vect{k}_s\left(\vect{q}\right)-\vect{k}_0$:
\begin{equation}
\vect{Q}=\left[q_x,q_y,k\left\{1-
\sqrt{1-\left(\frac{q}{k}\right)^2}
\right\}\right].
\label{eq_vettore_onda}
\end{equation}

If the schlieren blade is removed, both the waves with wave vector
$\vect{k}_s^+\left(\vect{q}\right)=\left[q_x,q_y,k_z\right]$ and
$\vect{k}_s^-\left(\vect{q}\right)=\left[-q_x,-q_y,k_z\right]$
generate fringes with wave vector
$\left[q_x,q_y\right]$. This leads to the shadowgraph method, where 
the power spectrum of the image, $S_{\delta
  I}\left(\vect{q}\right)$, shows deep modulations, and even vanishes
for some values of $q$ \cite{cannell95,cannell96}. This is due to the
fact that the modulations, generated by the interference of
the waves scattered at symmetric angles with the main beam, sum up
in counterphase. This effect is always present
at small wave vectors also for the heterodyne NFS
\cite{brogioli02}. In SNFS the zeroes are eliminated.

In order to fully appreciate the potentialities of the method we must
remind that NFS techniques are based on the fact that each point of
the CCD sensor is hit by light scattered at each scattering angle.
This imposes some constraints on the transverse dimension $d$ of the
scattering volume and its distance $Z$ from the sensor.
%
In the actual setup of fig.~\ref{fig_setup}, $Z$ represents the
distance of the sample from the plane, close to the sample,
which is imaged by the lens onto the CCD detector.
%
Each point of
the sensor collects light scattered at angles
$\vartheta<\vartheta_{max}$, where 
$\vartheta_{max}=d/Z$. The condition $Z\le d/\vartheta_{max}$
represents the Near Field Condition (NFC) \cite{giglio00,giglio01,brogioli02}.

In principle a close distance between the sample and the sensor
guarantees the access to a wide range of scattering angles without
having to use samples with large spatial extension. However, as we
will show shortly, this condition cannot be met for all the NFS
techniques, while for SNFS the distance $Z$ can be pushed to zero.
%
When $Z=0$ the sample is imaged onto the CCD detector.
%
Indeed, other NFS techniques impose a requirement on the minimal distance
between cell and sensor. As an example, in the HNFS technique this is
due to the fact that waves scattered at wave vectors $\vect{k}^+_s$
and $\vect{k}^-_s$
generate fringes with the same spacing when they interfere with the
transmitted beam onto the sensor. When $Z=0$, the phase difference of
the two waves is $\pi$, irrespective of the wavevector. Therefore, the
superposition of the two interference pattern shifted by $\pi$ gives rise
to a flat intensity distribution at $Z=0$. In order for the fringes to
become visible one can either block one of the scattered waves, as in
the schlieren technique described in this letter, or let the beam
propagate, so that a $Z$ dependent phase shift is introduced. This
second solution gives rise to the shadowgraph technique, where the
contrast of the fringes is modulated by the distance $Z$
\cite{cannell95,cannell96}. When $Z$ is large enough to guarantee
that the waves scattered at $\vect{k}^+_s$
and $\vect{k}^-_s$ come from non-overlapping
regions of the sample, the contrast of fringes ceases to depend on $Z$
and $q$, because the phase difference between the scattered waves
becomes random. This occurs when the diffraction aperture of the
sensor $Z\lambda/L$ onto the 
sample becomes larger than the sensor size $L$, $Z>L^2/\lambda$
\cite{brogioli02}.
This condition is not required by the SNFS technique, where arbitrarily
small distances and sample sizes can be used, provided that the NFC is
satisfied. This shows the great potentialities of SNFS, where by using
distances of the order of a few millimeters, a very compact instrument
can be obtained. Such a small setup could be very useful in many
applications where size is an issue, such as experiments under
microgravity conditions, medical apparatus, and in-situ monitoring in
general. 

We turn now to the non equilibrium fluctuations data. Non equilibrium
fluctuations have been investigated by different experimental
techniques, and the results so far seem to
confirm the theoretical expectations (see ref.~\cite{li94}).
The fluctuations arise whenever there is a
macroscopic gradient across the system. A simple way to build up a
concentration gradient is to carefully layer two miscible liquids one on
top of the other, as it is customarily done in a classic free diffusion
experiment.

Let us briefly remind the theoretical expectations for the scattered
intensity distribution from a layer of concentration gradient $\nabla c$.
The power spectra show a fast $q^{-4}$ power law decay at
long wave vectors, and a saturation at a constant value at small wave
vectors. The coupling of
velocity fluctuations with concentration fluctuations accounts for the
$q^{-4}$ decay of the power spectrum at the longer wave vectors.
Gravity is responsible for the
damping of small wave vector fluctuations.

The roll-off wave vector $q_{ro}$ where the transition occurs is:
\begin{equation}
q_{\mathrm{ro}}\left(t\right)=\left[
\frac{\beta g \nabla c\left(t\right)}{\nu D}
\right]^{1/4},
\end{equation}
where $g$ is the gravitational acceleration, $\nu$ is the kinematic viscosity, $D$ the is the diffusion
coefficient and $\nabla c$ is the largest concentration gradient along
the cell at time $t$.

For wavevectors smaller than $q_{ro}$, the scattered intensity
saturates to
\begin{equation}
\frac{I_s\left(q\to 0\right)}{I_0}=
\frac{K_B T}{8 \pi^2 \rho \beta g}
k^4\left(\frac{\partial n}{\partial c} \right)^2
\Delta c,
\label{eq_i_sat}
\end{equation}
where $\rho$ is the mass density,
$\beta=\rho^{-1}\left(\partial \rho/\partial c\right)_{p,T}$
is the solutal expansion coefficient and $I_0$ is the intensity of the
main beam.

\begin{figure}
\onefigure{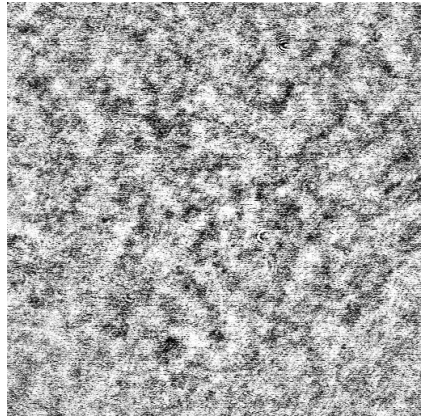}
\caption{
Image of the non-equilibrium concentration fluctuations during
the free diffusion of urea in water. The image was recorded
10 minutes after the start of the diffusion process. The side of the
image is 4\un{mm} in real space.
}
\label{fig_fluct}
\end{figure}

\begin{figure}
\onefigure{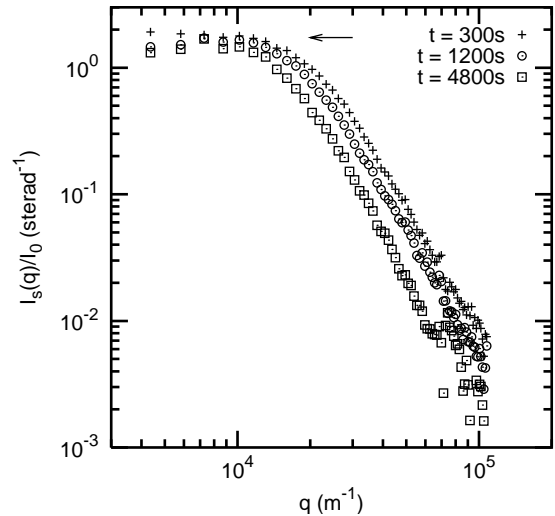}
\caption{
Relative scattered intensity due
to the non-equilibrium fluctuations in the free diffusion process.
}
\label{fig_grafico}
\end{figure}

The measurements have been taken on a water-urea sample, the same system
used in ref~\cite{brogioli00,brogioli00_2}, where a shadowgraph
technique was used. We report in fig.~\ref{fig_fluct} a 
%
schlieren
%
image of the fluctuations and in fig.~\ref{fig_grafico} the power
spectra obtained at different times during the free diffusion process.
Each curve was obtained by taking the average over typically a hundred
images taken over two hundred seconds, and static stray light was accounted
for by subtracting from each image the average intensity, as done for the
heterodyne method \cite{brogioli02}.
The potentiality of the method can be best appreciated by comparing the
curves in fig.~\ref{fig_grafico} with the ones shown in
fig.~3 of ref.~\cite{brogioli00} and obtained with a shadowgraph method.
The improvement  is very obvious. The wave vector range for the
shadowgraph data
is well below a decade, while for the schlieren data it covers a factor of
forty between $q_{min}$ and $q_{max}$. The shorter range for the shadowgraph method is
the result of the confinement between the low-$q$  wide instrumental zero,
and the tight sequence of zeros at larger wave vectors. No such a limitation
exists for the schlieren method where the instrumental transfer function is
flat. In spite of the moderate dynamic range (two decades), the
shadowgraph data appear noisy. At variance, as it
can be easily appreciated from fig.~\ref{fig_grafico}, the schlieren data show little noise
over a range of almost one thousand in the scattered intensity. 
%
Incidentally, this method profits on the huge number of pixels, leading
to a statistical accuracy well beyond the intrinsic 8-bit dynamic
range of the frame grabber.
%
The overall improved quality of the data permits a fairly good
estimate of the roll-off wave vector.
As time goes
on, the concentration gradient $\nabla c$ decreases as $t^{-1/2}$, and,
consequently, $q_{\mathrm{ro}} \propto t^{-1/8}$. Consistently with this, when
the time $t$ is increased by $16$ times from the first measurement to
the last, $q_{\mathrm{ro}}$ is reduced by about a factor $1.4\approx
16^{1/8}$.

Finally eq.~(\ref{eq_spettro}) can be used to
calculate with no adjustable parameters the absolute value of the
scattered intensity  at $q=0$, as from eq.~(\ref{eq_i_sat}).
An arrow in fig.~\ref{fig_grafico} indicate the no adjustable
parameter estimate for $I\left(q\to 0\right)$. As it
can be noticed, the agreement is very good. 

In conclusion, we feel that the schlieren method described in the present
work could be of great interest in performing extremely low angle light
scattering experiments that are of interest for large scale aggregation
processes, nucleation and growth phenomena on length scales that will merge
very naturally into regimes that traditionally are covered by more
classical methods, like interferometry.
The compactness of the layout may prove an asset for tight space
requiremens like microgravity setups, where the larger size objects can be
conveniently investigated without disturbing sedimentation processes.   

\acknowledgments

This work has been partially supported by the Italian Space Agency (ASI).

\end{document}